\documentclass[aps,prc,superscriptaddress,showpacs,amsmath,amssymb]{revtex4-1}
\usepackage{longtable,array}
\usepackage{graphicx}

\usepackage{bm}

\begin{document}

% ----------   NEW COMMANDS:  ------------
\newcommand{\half}{\mbox{$\textstyle \frac{1}{2}$}}
\newcommand{\ket}[1]{\left | \, #1 \right \rangle}
\newcommand{\bra}[1]{\left \langle #1 \, \right |}
\newcommand{\beq}{\begin{equation}}
\newcommand{\eeq}{\end{equation}}
\newcommand{\bea}{\begin{eqnarray}}
\newcommand{\eea}{\end{eqnarray}}
\newcommand{\req}[1]{Eq.\ (\ref{#1})}
\newcommand{\gcc}{{\rm~g\,cm}^{-3}}
\newcommand{\Compton}{\lambda\hspace{-.44em}\raisebox{.6ex}{\mbox{-$\!$-}}%
\raisebox{-.3ex}{}_{\hspace{-1pt}{\mbox{$_\mathrm{C}$}}}}
\newcommand{\kB}{k_\mathrm{B}}
\newcommand{\omc}{\omega_\mathrm{c}}
\newcommand{\omg}{\omega_\mathrm{g}}
\newcommand{\mel}{m_e}
\newcommand{\xr}{x_{\rm r}}
\newcommand{\EF}{\epsilon_{\rm F}}
\newcommand{\Ne}{{\cal N}_B(\epsilon)}
\newcommand{\Necl}{{\cal N}_0(\epsilon)}
\newcommand{\dfde}{{\partial f^{(0)} \over\partial\epsilon}}
\newcommand{\am}{a_\mathrm{m}}
\newcommand{\dd}{{\rm\,d}}
\newcommand{\vB}{\bm{B}}
\newcommand{\dotZ}{\mbox{$\dot{\mbox{Z}}$}}
\newcommand{\msun}{\mbox{$M_\odot$}}

\title{Large collection of astrophysical S-factors and its compact representation}

\author{A.\ V.\ Afanasjev}
\affiliation{Department of Physics and Astronomy, Mississippi State
University, P.O. Box 5167, Mississippi State, Mississippi 39762 USA}

\author{M.\ Beard}
\affiliation{Department of Physics \& The Joint Institute for Nuclear
Astrophysics, University of Notre Dame, Notre Dame, Indiana 46556,
USA}

\author{A.\ I.\ Chugunov}
\affiliation{Ioffe Physical-Technical Institute, Politekhnicheskaya
26, 194021 St.~Petersburg, Russia}

\author{M.\ Wiescher}
\affiliation{Department of Physics \& The Joint Institute for Nuclear
Astrophysics, University of Notre Dame, Notre Dame, Indiana 46556,
USA}

\author{D.\ G. Yakovlev}
\affiliation{Ioffe Physical-Technical Institute, Politekhnicheskaya
26, 194021 St.~Petersburg, Russia}
\affiliation{St.~Petersburg State Polytechnical University,
Politekhnicheskaya 29, St.~Petersburg 195251, Russia}
\affiliation{Department of Physics \& The Joint Institute for Nuclear
Astrophysics, University of Notre Dame, Notre Dame, Indiana 46556,
USA}

%\maketitle
\date{\today}
\begin{abstract}
Numerous nuclear reactions in the crust of accreting neutron stars
are strongly affected by dense plasma environment. Simulations of
superbursts, deep crustal heating and other nuclear burning phenomena
in neutron stars require astrophysical $S$-factors for these
reactions (as a function of center-of-mass energy $E$ of colliding
nuclei). A large database of $S$-factors is created for about 5,000
non-resonant fusion reactions involving stable and unstable isotopes
of Be, B, C, N, O, F, Ne, Na, Mg, and Si. It extends the previous
database of about 1,000 reactions involving isotopes of C, O, Ne, and
Mg. The calculations are performed using the S\~ao Paulo potential
and the barrier penetration formalism. All calculated $S$-data are
parameterized by an analytic model for $S(E)$ proposed before
[Phys.~Rev.~C~82, 044609 (2010)] and further elaborated here. For a
given reaction, the present $S(E)$-model contains three parameters.
These parameters are easily interpolated along reactions involving
isotopes of the same elements with only seven input parameters,
giving an ultracompact, accurate, simple, and uniform database. The
$S(E)$ approximation can also be used to estimate theoretical
uncertainties of $S(E)$ and nuclear reaction rates in dense matter,
as illustrated for the case of the $^{34}$Ne+$^{34}$Ne reaction in
the inner crust of an accreting neutron star.
\end{abstract}


\maketitle

%%%%%%%%%%%%%%%%%%%%%%
\section{Introduction}
\label{s:introduct}
%%%%%%%%%%%%%%%%%%%%%%

Nuclear burning is an important ingredient of stellar structure and
evolution \cite{bbfh57,fh64,clayton83}. Simulating the evolution and
observational manifestations of different stars (from main-sequence,
to giants and supergiants, presupernovae, white dwarfs and neutron
stars), requires the rates of many reactions, involving different
nuclei -- light and heavy, stable and neutron-rich. These rates are
derived from respective reaction cross sections $\sigma(E)$, where
$E$ is the center-of-mass energy of reactants.

The rates of many reactions which occur in the classical
thermonuclear regime -- for instance, in main-sequence stars -- are
well known. However nuclear burning in dense plasma of white dwarfs
and neutron stars \cite{st83}, which affects the evolution and many
observational manifestations of these objects, may proceed in other
regimes, under strong effects of plasma screening and pycnonuclear
tunneling through Coulomb barrier (e.g., Refs.\
\cite{svh69,yak2006,cd09}). The burning powers nuclear explosions in
surface layers of accreting white dwarfs (nova events), in cores of
massive accreting white dwarfs or in binary white dwarf mergers (type
Ia supernovae) \cite{NiWo97,hoeflich06,vankerkwijketal2010}, and in
surface layers of accreting neutron stars (type I X-ray bursts and
superbursts; e.g., Refs.\
\cite{sb06,schatz03,cummingetal05,Brown06,cooperetal09}). Nova events
and type I X-ray bursts are mostly driven by the proton capture
reactions of the hot CNO cycles and by the rp-process. This burning
is thermonuclear, without any strong effects of dense plasma
environment. Type Ia supernovae and superbursts are driven by the
burning of carbon, oxygen, and heavier elements
%(e.g.,\cite{sb06,schatz03,cummingetal05,Brown06})
at high densities, where the plasma screening effects can be
substantial. It is likely that pycnonuclear burning of neutron-rich
nuclei (e.g., $^{34}$Ne+$^{34}$Ne) in the inner crust of accreting
neutron stars in X-ray transients (in binaries with low-mass
companions; e.g.\ Refs.\ \cite{hz90,hz03,Brown06}) provides an
internal heat source for these stars. If so, it powers \cite{Brown98}
thermal surface X-ray emission of neutron stars observed in quiescent
states of X-ray transients (see, e.g., Refs.\
\cite{Brown06,pgw06,lh07}) although other energy sources can also be
important there (e.g., Ref.\ \cite{bk09}).

Therefore, in dense plasma of evolved stars, standard classical
thermonuclear reaction rates may be unavailable and/or inapplicable.
Accordingly, one needs to construct many reaction rates, valid over
all burning regimes, starting from reaction cross sections
$\sigma(E)$. Here we focus on non-resonant fusion reactions involving
$(A_1,Z_1)$ and $(A_2,Z_2)$ reactants ($A_i$ and $Z_i$ stand for
their mass and charge numbers), which may occur in evolved stars. The
cross section $\sigma(E)$ is conveniently expressed through the
astrophysical $S$-factor,
\begin{equation}
   \sigma(E) = E^{-1}\, \exp(-2 \pi \eta)\,S(E),
\label{e:sigma}
\end{equation}
where $\eta= \alpha/(\hbar v)=\sqrt{E_R/E}$ is the Sommerfeld
parameter, $v=\sqrt{2E/ \mu}$ is the relative velocity of the
reactants at large separations, $\alpha=Z_1 Z_2 e^2$, $E_R=\alpha^2
\mu/(2 \hbar^2)$ is analogous to the Rydberg energy in atomic
physics, and $\mu$ is the reduced mass. The factor $\exp(-2 \pi
\eta)$ is proportional to the probability of penetration through the
pure Coulomb barrier $U(r)=\alpha/r$ with zero angular orbital
momentum, assuming that this pure Coulomb barrier extends to $r \to
0$ (as for point-like nuclei); $E^{-1}$ factorizes out the well-known
pre-exponential low-energy dependence of $\sigma(E)$. The advantage
of this representation is that $S(E)$ is a more slowly varying
function of $E$ than $\sigma(E)$.

For astrophysical applications, one needs $S(E)$ at low energies, $E
\lesssim $ a few MeV. Even for beta-stable nuclei experimental
measurements of $\sigma(E)$ at such energies are mainly not available
because the Coulomb barrier becomes extremely thick making
$\sigma(E)$ exponentially small. It will be even more difficult to
get such data for neutron-rich nuclei. Moreover, the modelling of the
processes at different astrophysical sites (such as dynamic reaction
network modeling of the neutron star crust composition  at high
densities \cite{MW-08}) requires the knowledge of $S$-factors for a
large variety of reactions, many of which involve neutron-rich
nuclei. The experimental study of so many reactions is definitely
beyond existing and near-future capabilities. Therefore, one must
rely on theoretical calculations.

%The huge number of reactions %involved in these processes
%also requires
%the simplification for the calculations and storage of the
%that the calculation and storage of the $S$-factors be simplified.
Previously \cite{paper1} we calculated $S(E)$ for 946 reactions
involving isotopes of C, O, Ne, and Mg. Also, we proposed
\cite{paper2} a simple analytic model for $S(E)$ and used it to fit
calculated $S$-factors. %Our aim here is to significantly enlarge our
%database by calculating $S(E)$ for new reactions involving isotopes
%of Be, B, C, N, O, F, Ne, Na, Mg, and Si.
The present paper significantly enlarges the database. We have
extended the calculations to incorporate new reactions between
even-even and odd($Z$)-even isotopes. In particular, we have
calculated $S(E)$ between isotopes of  Be, B, C, N, O, F, Ne, Na,
Mg, and Si. Our new database of $S(E)$ contains 4,851 reactions.
Moreover, we elaborate and simplify the analytic $S(E)$-model, and
use it to parameterize all $S$-factors with the minimum number of fit
parameters producing thus an ultracompact and uniform database
convenient in applications.

%%%%%%%%%%%%%%%%%%%%%%%%%%%%%%%%%%%%%%%%%%%%%%%%%%%%%%%%%%%%%%%%%%%%
\section{Calculations}
\label{s:calculations}
%
%%%%%%%%%%%%%%%%%%%%%%%%%%%%%%%%%%%%%%%%%%%%%%%%%%%%%%%%%%%%%%%%%%%%

Consider a set of $S$-factors for non-resonant fusion reactions
involving various isotopes of 10 elements, Be, B, C, N, O, F, Ne, Na,
Mg, and Si. The $S$-factors for the reactions involving even-even
isotopes of C, O, Ne, and Mg have been calculated recently in our
previous paper \cite{paper1}. Calculations for other reactions are
original and include even-even and odd($Z$)-even stable, proton-rich,
neutron-rich, and very neutron-rich isotopes. Such isotopes can
appear during nuclear burning in stellar matter, particularly, in the
cores of white dwarfs and envelopes of neutron stars. The
calculations have been performed using the S\~ao Paulo potential in
the frame of the barrier penetration model \cite{gas2005}. Nuclear
densities of reactants have been obtained in Relativitic
Hartree-Bogoliubov (RHB) theory \cite{VALR.05} employing the NL3
parametrization for relativistic mean field Lagrangian \cite{NL3} and
Gogny D1S force for pairing. The model is based on the standard
partial wave decomposition ($\ell=0,1,\ldots$) and considers the
motion of reacting nuclei in the effective potential discussed in
Sec.\ \ref{s:ell} [see Eq.\ (\ref{e:Vcentrifug}) there]. The
numerical scheme is parameter-free and relatively simple for
generating a set of data for many non-resonant reactions involving
different isotopes.

The reactions in question are listed in Table \ref{tab:reactions}.
All reactants are either even-even or odd-even nuclei. We consider 55
reaction types, such as Si+Si and Be+B, with the range of mass
numbers for both species given in the columns 2 and 3 of Table
\ref{tab:reactions}. For each reaction, $S(E)$ has been computed on a
dense grid of $E$ (with the energy step of 0.1 MeV) from 2~MeV to a
maximum value $E_\mathrm{max}$ (also given in Table
\ref{tab:reactions}) covering wide energy ranges below and above the
Coulomb barrier. The last column in Table \ref{tab:reactions}
presents the number of considered reactions.

$S(E)$-factors calculated using the S\~ao Paulo potential have been
compared previously \cite{yak2006, gas2005,saoPauloTool} with
experimental data and with theoretical calculations performed using
other models such as coupled-channels and fermionic molecular
dynamics ones. Let us stress that the calculated values of $S(E)$ are
uncertain due to nuclear physics effects -- due to using the S\~ao
Paulo model with the NL3 nucleon density distribution. As shown, for
instance, in Ref.\ \cite{saoPauloTool}, typical expected
uncertainties for the reactions involving stable nuclides are within
a factor of 2, with maximum up to a factor of 4. For the reactions
involving unstable nuclei, typical uncertainties can be as large as
one order of magnitude, reaching two orders of magnitude at low
energies for the reactions with very neutron-rich isotopes. These
uncertainties reflect the current state of the art in our knowledge
of $S(E)$.
%Since we will fit the calculated $S(E)$ with analytic
%expressions, very accurate fitting is not crucial.

%%%%%%%%%%%%%%%%%%%%%%%%%%%%%%%%%%%%%%%%%%%%%%%%%%%%%%%%%%%%%%%%%%%%%%%%%
\section{Analytic model}
\label{s:model}

%%%%%%%%%%%%%%%%%%%%%%%%%%%%%%%%%%%%%%%%%%%%%%%%%%%%%%%%%%%%%%%%%%%%%%%%%

%%%%%%%%%%%%%%%%%%%%%%%%%%%%%%%%%%%%%%%%%%%%%%%%%%%%%%%%%%%%%%%%%%%%%%%%%
\subsection{General remarks}
\label{s:general}

%%%%%%%%%%%%%%%%%%%%%%%%%%%%%%%%%%%%%%%%%%%%%%%%%%%%%%%%%%%%%%%%%%%%%%%%%

Let us recall the main features of the analytic model for $S(E)$
proposed in Ref.\ \cite{paper2} and used there to fit the initial set
of 946 $S$-factors. We will elaborate and simplify the model, and fit
much larger set of data.
%; the employed simplification will
%in reality lead to an improvement of the quality of the fit.

The model \cite{paper2} is based on semi-classical consideration of
quantum tunneling through an effective potential $U(r)$ (which is
purely Coulombic at large separations $r$ but is truncated by nuclear
interactions at small $r$ when colliding nuclei merge). The reaction
cross section at $E< E_C$ (where $E_C$ is the barrier height) is
taken in the form
\begin{eqnarray}
    \sigma(E)&=&\frac{S_0}{E} \exp \left[
      -{2 \over \hbar}
       \int_{r_1}^{r_2} {\rm d}r\, \sqrt{2 \mu ( U-E)} \right],
\label{e:belowsigma}
\end{eqnarray}
where $r_1$ and $r_2$ are classical turning points; $S_0$ is a slowly
varying function of $E$ treated as a constant; it has the same
dimension as  $S(E)$, but should not be confused with it.
%At $E>E_C$
%in the semi-classical approximation the Coulomb barrier is
%transparent and $\sigma(E) \propto 1/\sqrt{E}$.

To obtain tractable formulae for $S(E)$, in Ref.\ \cite{paper2} we
employed the natural and simplest approximation of $U(r)$,
\begin{eqnarray}
&&   U(r)={\alpha \over r}~~~\mathrm{at}~~r\ge R_{C1},
\nonumber \\
&&   U(r)=E_C \left[ 1 - \beta {(r-R_C)^2 \over R_C^2}
   \right]~~~\mathrm{at}~~r<R_{C1},
\label{e:U}
\end{eqnarray}
which is a pure Coulomb potential at $r \ge R_{C1}$ and an inverse
parabolic potential at smaller $r$ (see Fig.\ 1 in Ref.\
\cite{paper2}). The parabolic segment truncates the effective
interaction at small separations; $E_C=U(R_C)$ is the maximum of
$U(r)$, that is the barrier height.  We required $U(r)$ and its
derivative to be continuous at $r=R_{C1}$. Instead of $\beta$ we will
often use $\delta=(R_{C1}-R_C)/R_C$, which characterizes the width of
the peak maximum of $U(r)$. Then $U(r)$ is specified by two
parameters, $E_C$ and $\delta$, with \cite{paper2}
\begin{eqnarray}
  &&  R_C={ \alpha (2+3 \delta) \over 2 E_C (1+\delta)^2}, \quad
    \beta= {1 \over \delta (2+3 \delta)},
\nonumber \\
  &&  R_{C1}=R_C\,(1+\delta), \quad
    E_{C1}=U(R_{C1})=E_C\, {2+ 2 \delta \over 2 +3 \delta}.
\label{e:Upar}
\end{eqnarray}
The potential $U(r)$ passes through zero at $r=R_{C0}=R_C (1 -
\beta^{-1/2})$; its behavior at smaller $r$ is unimportant (in our
approximation). Realistic models should correspond to $\beta \gg 1$
(the small-$r$ slope of $U(r)$ should be sharp; $R_{C0}$ should be
positive) which translates into $\delta \ll {1 \over 3}$ (because
$\beta=1$ corresponds to $\delta={1 \over 3}$).

%%%%%%%%%%%%%%%%%%%%%%%%%%%%%%%%%%%%%%%%%%%%%%%%%%%%%%%%%%%%%%%%%%%%%%%%%
\subsection{Sub-barrier energies}
\label{s:lowE}

%%%%%%%%%%%%%%%%%%%%%%%%%%%%%%%%%%%%%%%%%%%%%%%%%%%%%%%%%%%%%%%%%%%%%%%%%

With the potential (\ref{e:U}) at $E<E_C$ one has
\begin{equation}
  S(E)=S_0\,\exp \Psi(E),
\label{e:S(E)}
\end{equation}
where $\Psi(E)$ is taken analytically and has a rather complicated
form given by Eq.~(9) in Ref.\ \cite{paper2}. Notice that in our
semi-classical approximation $\Psi(E_C)=2 \pi \eta_C$, where
$\eta_C=\eta(E_C)$ is the Sommerfeld parameter at $E=E_C$.

In this paper we propose a simplified expression for $S(E)$ at
$E<E_C$. Let us recall that $\Psi(E)$ at $E<E_C$ is accurately
approximated \cite{paper2} by the Taylor expansion
$\Psi(E)=g_0+g_1E+g_2E^2+\ldots$ The explicit expressions for the
expansion coefficients $g_0$, $g_1$, and $g_2$ in terms of $E_C$ and
$\delta$ are given by Eqs.\ (14) and (15) of Ref.\ \cite{paper2}. At
$E<E_C$ we suggest the approximation
\begin{eqnarray}
 && \Psi(E)= g_0+g_1E+g_2E^2,
\label{e:approxpsi}
 \\
 && g_2=(2\pi\eta_C-g_0-g_1 E_C)/E_C^2.
\label{e:g3}
\end{eqnarray}
The expressions for $g_0$ and $g_1$ will be taken from Ref.\
\cite{paper2}, while $g_2$ is introduced in such a way to satisfy the
condition $\Psi(E_C)=2\pi \eta_C$. Our polynomial approximation
(\ref{e:approxpsi}) is much simpler than the exact expression (9) of
Ref.\ \cite{paper2} but gives nearly the same accuracy in
approximating all $S(E)$-factors considered in Sec.\
\ref{s:calculations}. Nevertheless, it is still not very convenient
in applications because Eqs.\ (14) and (15) of Ref.\ \cite{paper2}
for $g_0$ and $g_1$ are rather complicated. Therefore we further
simplify these equations by adopting the limit of $\delta \ll 1$,
which is sufficient for applications in Sec.\ \ref{s:Fits}. In this
limit we have \cite{paper2}
\begin{eqnarray}
&& g_0=\sqrt{E_R \over E_C}\;(8-\pi\,\sqrt{2 \delta} -2 \delta),
\label{e:g0}\\
&& g_1=-\sqrt{E_R \over E_C^3}\,\left({4 \over 3}- \pi \sqrt{2
\delta} -\delta \right). \label{e:g1}
%
%&& g_2=- \sqrt{E_R \over E_C^5} \, \left({1 \over 5}-{\delta \over 4}
%\right).
%
%\label{e:g2}
\end{eqnarray}
%
%Equations (\ref{e:g0}) and (\ref{e:g1}) are presented in Ref.\
%\cite{paper2}; Eq.\ (\ref{e:g2}) is obtained in the same manner.
Thus, Eqs.\ (\ref{e:S(E)})--(\ref{e:g1}) give simple and practical
expressions for $S(E)$ at $E<E_C$. In our model, $S(E)$ is determined
by the three parameters, $E_C$, $\delta$, and $S_0$; $E_C$ and
$\delta$ determine the shape of the potential $U(r)$; $S_0$ specifies
the efficiency of fusion reaction.

%%%%%%%%%%%%%%%%%%%%%%%%%%%%%%%%%%%%%%%%%%%%%%%%%%%%
\subsection{Contribution of $\ell>0$ waves}
\label{s:ell}

The proposed model is phenomenological, being based on $s$-wave
semi-classical tunneling through a spherical potential barrier
$U(r)$. The actual reaction cross section $\sigma(E)$ contains
contribution of different $\ell$ waves, with $\ell=0,1,2,\ldots$ For
a given multipolarity $\ell$, we have a quantum mechanical scattering
problem of two nuclei moving in an effective potential
\begin{equation}
    V_\mathrm{eff}(r)=U(r)+V_\ell(r),\quad V_\ell(r)=\frac{\hbar^2 \ell(\ell+1)}{2\mu r^2},
\label{e:Vcentrifug}
\end{equation}
where $V_\ell(r)$ is the centrifugal potential. Then the cross
section is (e.g., Refs.\ \cite{vas81,leandro04})
\begin{equation}
    \sigma(E)={\pi \over k^2}\,\sum_{\ell=0}^\infty (2\ell+1)
    T_\ell(E)\,P_\ell(E),
\label{e:sigmaell}
\end{equation}
where $k$ is wave-number ($E=\hbar^2 k^2/2\mu$), $T_\ell(E)$ is the
transmission coefficient, and $P_\ell(E)$ is the fusion probability
for the penetrating wave. At the low energies of our interest, one
traditionally assumes $P_\ell(E)=1$.

Let us return to our $S(E)$-model at $E<E_C$. In accordance with
(\ref{e:sigmaell}), we have $S(E)=\sum_\ell S^{(\ell)}(E)$, where
$S^{(\ell)}(E)$ is an $\ell$-wave contribution to $S(E)$ to be
calculated using the effective potential $V_\mathrm{eff}(r)$. At
$E<E_C$, all $\ell$ waves refer to subbarrier motion, and the
transmission coefficient $T_\ell(E)$ decreases evidently with the
growth of $\ell$.  Then $S(E)$ can be written as
\begin{equation}
   S(E)=S^{(0)}(E) \,J(E), \quad J(E)=1+\sum_{\ell=1}^\infty
   (2\ell+1){T_\ell(E)\over T_0(E)}.
\label{e:J(E)}
\end{equation}
Here, $S^{(0)}(E)$ is the $s$-wave contribution, and the sum in
$J(E)$ is the correction due to  waves with $\ell \geq 1$.

A simple analysis shows that $S(E)$ is typically determined by
several lowest $\ell$ at $E<E_C$ for reactions considered in Sec.\
\ref{s:calculations}. The correction factor $J(E)$ appears to be
essentially higher than 1 (values $\ell>0$ are important) but is a
slowly varying function of $E$. In other words, the transmission
coefficients $T_\ell(E)$ at these $\ell$ are similar functions of
$E$. Their main energy dependence is the same as for $s$-wave,
$S^{(0)}(E)$. A crude estimate at $E$ noticeably below $E_C$ gives
$J(E)\sim 1+\sqrt{E_C/E_0}$, where $E_0=\hbar^2/(2 \mu R_C^2)$ is the
characteristic quantum of centrifugal energy (typically, $E_0 \ll
E_C$). To simplify the model, at $E<E_C$ we suggest the approximation
\begin{eqnarray}
   S(E)&=&S_{0s}J_0\,\exp(g_0+g_1E+g_2E^2),
\label{e:SElowerEc}\\
    J_0 &=& 1+j_0\sqrt{E_C / E_0}.
\label{e:j}
\end{eqnarray}
Here, $S_{0s}$ is the $s$-wave contribution to $S_0$, $J(E)$ is
approximated by energy-independent constant $J_0$ (that is specified
by a constant $j_0$); $J_0$ (or $j_0$) can be treated as a parameter
which characterizes the importance of higher multipolarities
$\ell>0$. One often states in quantum mechanics that the main
contribution to scattering at low $E$ comes from $\ell=0$. This
statement does not apply to the potential considered here, which has
strong attraction at $r=0$. In this case, higher $\ell$ are important
even at very low $E$.

Finally, at $E=E_C$, the $s$-wave barrier is removed and  we have
$T_0(E_C)=1$  in Eq.~(\ref{e:sigmaell}). Then the $s$-wave partial
cross section is $\sigma_0(E_C)=\pi/k_C^2$, where $k_C$ is the
wave-number referring to $E=E_C$. From this $\sigma_0(E_C)$ and
Eq.~(\ref{e:sigma}) we immediately get:
\begin{equation}
    S_{0s}={\pi \hbar^2 \over 2 \mu}= 0.6566\, {A_1+A_2 \over
    A_1A_2}~~\mathrm{MeV~barn}.
\label{e:S0s}
\end{equation}
Therefore, $S_{0s}$ is specified by $A_1$ and $A_2$. The
astrophysical $S$-factor at $E<E_C$ is determined by the parameters
$E_C$ and $\beta$ (or $\delta$) of the $U(r)$ potential and by the
factor $J_0$ (or $j_0$).

%%%%%%%%%%%%%%%%%%%%%%%%%%%%%%%%%%%%%%%%%%%%%%%%%%%%%%%%%%%%%%%%%%%%%%%%%
\subsection{Above-barrier energies}
\label{s:highE}
%%%%%%%%%%%%%%%%%%%%%%%%%%%%%%%%%%%%%%%%%%%%%%%%%%%%%%%%%%%%%%%%%%%%%%%%%

At $E>E_C$ the effective barrier is transparent for some low-$\ell$
waves. In this case we adopt the simplest barrier-penetration model
with $T_\ell(E)=1$ if the  $V_\mathrm{eff}(r)$ barrier  is
transparent at a given $E$, and with $T_\ell(E)=0$ otherwise. The
cross section is then given by Eq.\ (\ref{e:sigmaell}) with
$T_\ell=1$ and $P_\ell=1$, where the sum is taken from $\ell=0$ to
some maximum $\ell_0(E)$ at which $V_\mathrm{eff}(r)$ becomes
classically forbidden. To be transparent at lower $\ell$, the
$V_\mathrm{eff}(r)$ potential should have a pocket (with a local
minimum) and a barrier (with a local peak) at $r<R_C$; let
$r=r_0<R_C$ be the peak point. It is well known that in this case at
$E>E_C$ the cross section becomes (e.g., Ref.\ \cite{vas81})
\begin{equation}
   \sigma(E)={\pi \over k^2}\,(\ell_0(E)+1)^2.
\label{e:above_barrier}
\end{equation}

With increasing $E$, the range of $\ell \leq \ell_0(E)$ widens. In
the spirit of semi-classical approximation, at $E>E_C$ we can treat
$\ell_0(E)$ as a continuous variable. As long as $E$ is not too much
higher than $E_C$, the local peak point $r_0$ of $V_\mathrm{eff}(r)$
is close to $R_C$. To locate this peak, it is convenient to introduce
$x=(R_C-r)/R_C$ instead of $r$ and linearize $V_\ell(r)$ in terms of
$x$, keeping constant and linear terms. The peak occurs at
$x=x_0=(\ell+1)\ell E_0/(E_C\beta)$ at which ${\rm
d}V_\mathrm{eff}/{\rm d}x=0$. The peak height is
$V_\mathrm{eff}(r_0)=E_C+(\ell+1)\ell E_0+(\ell+1)^2 \ell^2
E_0^2/(E_C\beta)$. At a given $E$ the barrier becomes classically
forbidden when $V_\mathrm{eff}(r_0)=E$, which gives a quadratic
equation for $(\ell_0+1)\ell_0$. Solving this equation, we have
\begin{equation}
  (\ell_0+1)\ell_0=
  \frac{\sqrt{E_C^2\beta^2+4E_C \beta (E-E_C)}-E_C \beta}{2E_0}.
\label{e:l0}
\end{equation}
Now we can easily find $\ell_0(E)$. Substituting it into
Eq.~(\ref{e:above_barrier}) gives a closed expression for $\sigma(E)$
at $E>E_C$.

However, such a solution seems overcomplicated. It can be further
simplified if we notice that at not too high $E$ we typically have
$E-E_C \ll \beta E_C$ and expand the expression containing square
root. This gives
\begin{equation}
    y(E)\equiv(\ell_0+1)\ell_0=
  \frac{E-E_C}{E_0}\,\left( 1-  \frac{E-E_C}{\beta E_C}  \right),
\label{e:l01}
\end{equation}
the second term in the parentheses being a small correction.

One has $\ell_0(E)\gg 1$ very soon after $E$ exceeds the barrier
$E_C$; then $(\ell_0+1)\ell_0 \approx \ell_0^2$. On the other hand,
at $E=E_C$ in the adopted approximation we should have
$\ell_0(E_C)=0$ and $\sigma_0(E_C)=\pi/k_C^2$. Naturally, our
approximation of continuous $\ell_0(E)$ is inaccurate just near the
threshold ($E \to E_C$), where $s$-wave nuclear collisions proceed
above the threshold, while higher-$\ell$ collisions operate either in the
subbarrier regime or in the transition regime (emerging from under
respective barriers with growing $E$). It is a complicated task to
describe accurately $\sigma(E)$ and $S(E)$ at $E \approx E_C$. In
order to preserve the simplicity of our $S(E)$ model we propose
writing the cross section $\sigma(E)$ at $E>E_C$ as
\begin{equation}
   \sigma(E)={\pi \over k^2}\,\sqrt{y^2(E)+J_0^2},
\label{e:sigma_above_barrier}
\end{equation}
where $y(E)$ is given by Eq.\ (\ref{e:l01}) and $J_0$ by Eq.\
(\ref{e:j}). Then at $E=E_C$ this equation matches the cross section
model proposed for subbarrier energies (Sec.\ \ref{s:lowE}) while at
$E-E_C \gtrsim J_0 E_0$ it reproduces the semi-classical cross
section (\ref{e:above_barrier}).

Translating the cross section (\ref{e:sigma_above_barrier}) into
$S(E)$, at $E>E_C$ we finally have
\begin{equation}
   S(E)=S_{0s}\,\exp(2\pi\eta)\,\sqrt{y^2(E)+J_0^2}.
\label{e:S_above_Ec}
\end{equation}
At $E-E_C\gg J_0E_0$ the leading term in the astrophysical $S$-factor
is $S(E)\approx S_{0s}\,\exp(2 \pi \eta)\,(E-E_C)/E_0$. In the
previous work we described $S(E)$ by a phenomenological expression
(Eq.~(6) in Ref.\ \cite{paper2}). It reproduces this leading term at
$J_0 E_0 \ll E-E_C\ll E_C$ (see the term containing $\xi$ in Eq.\ (6)
of Ref.\ \cite{paper2}) but diverges from it at higher $E$. Although
we do not intend to propose an accurate approximation of $S(E)$ at
high energies (few barrier energies $E_C$ and higher), we remark that
the phenomenological expression (6) for $S(E)$ at $E>E_C$ in Ref.\
\cite{paper2} is now replaced by Eq.\ (\ref{e:S_above_Ec}) which
reproduces the correct semi-classical behavior of $S(E)$ above the
barrier. Our new expression (\ref{e:S_above_Ec}) is self-sufficient,
it contains no extra input parameters in addition to those introduced
for describing $S(E)$ at $E<E_C$ in Sec.\ \ref{s:lowE}.

%%%%%%%%%%%%%%%%%%%%%%%%%%%%%%%%%%%%%%%%%%%%%%%%%%%%%%
\subsection{How to calculate $S(E)$}
\label{s:instruction}

%For convenience of users we outline the procedure of calculating
%$S(E)$ in our model.
For any fusion reaction, $S(E)$ is determined by
three input parameters, $E_C$, $\beta$ (or $\delta$, see Eq.\
(\ref{e:Upar})) and $J_0$ (or $j_0$, see Eq.\ (\ref{e:j})). The first
two parameters specify the shape of the effective potential $U(r)$,
Eq.\ (\ref{e:U}); $J_0$ (or $j_0$) takes into account the
contribution of higher $\ell=1,2,\ldots$ to $S(E)$. At $E<E_C$ the
astrophysical factor $S(E)$ is given by Eq.\ (\ref{e:SElowerEc}), and
at $E\geq E_C$ it is given by (\ref{e:S_above_Ec}). The factor
$S_{0s}$ entering these equations is defined by (\ref{e:S0s}). The
parameters $g_0$, $g_1$, and $g_2$ in Eq.\ (\ref{e:SElowerEc}) are
given by (\ref{e:g0}), (\ref{e:g1}), and (\ref{e:g3}).

By construction, our $S(E)$ model can be accurate at $E$ up to a few
$E_C$. Although the model is generally based on first principles, it
is phenomenological in a narrow energy range near the barrier
($|E-E_C|\lesssim J_0 E_0$). Our current $S(E)$ model is simpler than
its previous version \cite{paper2} and contains 3 input parameters
instead of 4 in \cite{paper2}. We will see that the current model is
more accurate than the previous one.

Let us stress that our new version implies $\delta \ll 1$ (actually,
$\delta \lesssim 0.1$) meaning that the maximum of the effective
potential $U(r)$ [see Eq.\ (\ref{e:U})] is rather sharp. For broader
maximum ($0.1 \lesssim \delta \leq 1/3$), it would be more
appropriate to use the same Eq.~(\ref{e:S_above_Ec}) at $E \geq E_C$
but replace our Eq.~(\ref{e:SElowerEc}) at $E<E_C$ by Eqs.\ (5) and
(9) of Ref.\ \cite{paper2}, setting $S_0=S_{0s}J_0$ in (9). Although
this would complicate the expression for $S(E)$ at $E<E_C$, the fit
parameters would preserve their meaning.

%%%%%%%%%%%%%%%%%%%%%%%%%%%%%%%%%%%%%%%%%%%%%%%%%%%%%%%%%%%%%%%%%%
\section{Fits}
\label{s:Fits}
%
%%%%%%%%%%%%%%%%%%%%%%%%%%%%%%%%%%%%%%%%%%%%%%%%%%%%%%%%%%%%%%%%%%

Let us approximate all calculated $S(E)$ using the analytic model of
Sec.\ \ref{s:model}. In Ref.\ \cite{paper2} we approximated $S(E)$
for the 946 reactions involving C, O, Ne, and Mg isotopes with the
first version of the model. That approximation is sufficiently
accurate, but we approximate those data again together with the new
data using the elaborated $S(E)$-model to obtain a uniform database
with minimum number of input parameters.

We consider reactions of each type (each line in Table
\ref{tab:reactions}) separately, and apply the analytic model
(\ref{e:SElowerEc}) and (\ref{e:S_above_Ec}) to every reaction. In
this manner we determine 3 fit parameters, $E_C$, $\delta$ and $j_0$,
for every reaction. For instance, we have $3 \times 120=360$
parameters for Si+Si reactions. However, we notice that we can set
$\delta$ and $j_0$ constant for all reactions of a given type (for
example, $\delta=0.0409$ and $j_0=2.8162$ for all Si+Si reactions)
without greatly increasing the fit errors. Such constant $\delta$ and
$j_0$ are given in Table \ref{tab:fit}. Notice that while $j_0$ is
constant for a given reaction type, $J_0$ is given by the scaling
relation (\ref{e:j}) and differs from one reaction to another.

Still, we need to specify the barrier height $E_C$ for every
reaction. Collecting the values of $E_C$ for all reactions of each
type, we were able to fit them by the same analytic expression as in
Ref.\ \cite{paper2}:
\begin{eqnarray}
      E_C &= & {\alpha / R_{12}}, \quad R_{12}=R+
       \Delta R_1 \, |A_1-A_{10}|+\Delta R_2 \, |A_2-A_{20}|,
\label{e:ECfit}
\end{eqnarray}
where $A_{10}=2Z_1$ and $A_{20}=2Z_2$ are mass numbers of most
stable isotopes;
$\Delta R_1=\Delta R_{1a}$ at $A_1 \geq A_{10}$;
$\Delta R_1=\Delta R_{1b}$ at $A_1 < A_{10}$;
$\Delta R_2=\Delta R_{2a}$ at $A_2 \geq A_{20}$;
$\Delta R_2=\Delta R_{2b}$ at $A_2 < A_{20}$.
This gives 5 new fit parameters $R$, $\Delta R_{1a}$, $\Delta
R_{2a}$, $\Delta R_{1b}$, $\Delta R_{2b}$ (also given in Table
\ref{tab:fit}) for each reaction type, and, hence, 7 parameters in
total.
%Naturally, we have $\Delta R_2=\Delta R_1$ for the reactions
%involving isotopes of the same element (e.g., Si+Si).

As in Ref.\ \cite{paper2}, our fit procedure is based on standard
relative deviations of calculated (calc) and fitted (fit)
$S(E)$-factors. The absolute value of such a deviation in a point $E$
is $\eta(E)=|1-S_\mathrm{fit}(E)/S_\mathrm{calc}(E)|$. Fitting has
been done by minimizing root-mean-square (rms) deviation
$\eta_\mathrm{rms}$ over all energy grid points for all reactions
involved in a fit. Column 9 of Table \ref{tab:fit} lists
$\eta_\mathrm{rms}$ for all reactions of a given type over all energy
grid points (e.g., over $120 \times 379= 45,480$ points for the Si+Si
reactions). Column 10 presents the maximum absolute value of the
relative standard deviations, $\eta_\mathrm{max}$, over all these
reactions and points. Root-mean-square values $\eta_\mathrm{rms}$ are
reasonably small; they vary from $\eta_\mathrm{rms}\approx 0.07$ for
B+B, B+C, and B+N reactions to $\eta_\mathrm{rms}\approx 0.28$ for
Si+Si reactions. Maximum relative deviations are larger, reaching
$\eta_\mathrm{max}\approx 1.12$ for Si+Si. Such large values of
$\eta_\mathrm{max}$ hide the proper maximum difference of
$S_\mathrm{fit}(E)$ to $S_\mathrm{calc}(E)$ in a fit.

To visualize this difference, in column 11 we list the maximum value
of the parameter $\widetilde{\eta}(E)$ that we define as
$\widetilde{\eta}(E)=S_\mathrm{fit}(E)/S_\mathrm{calc}(E)-1$ for
$S_\mathrm{fit}(E)>S_\mathrm{calc}(E)$ and as
$\widetilde{\eta}(E)=S_\mathrm{calc}(E)/S_\mathrm{fit}(E)-1$ for
$S_\mathrm{fit}(E)\leq S_\mathrm{calc}(E)$. Thus defined, we have
$\widetilde{\eta}=\eta$ at $S_\mathrm{fit}>S_\mathrm{calc}$ but
$\widetilde{\eta}>\eta$ at $S_\mathrm{fit}<S_\mathrm{calc}$. For
$\eta \ll 1 $ we always have $\widetilde \eta \approx \eta$, but for
$\eta \gtrsim 1$ the value of $\widetilde{\eta}$ can be much larger
than $\eta$. One can see that $\widetilde{\eta}_\mathrm{max}+1$ is
the maximum value among ratios $S_\mathrm{fit}/S_\mathrm{calc}$ (at
$S_\mathrm{fit}>S_\mathrm{calc}$) and
$S_\mathrm{calc}/S_\mathrm{fit}$ (at $S_\mathrm{fit}\leq
S_\mathrm{calc}$) in a fit sample. For our Si+Si reactions we have
$\widetilde{\eta}_\mathrm{max}+1\approx 5.40$. This relatively large
difference between $S_\mathrm{calc}$ and $S_\mathrm{fit}$ occurs at
one energy point in one of the 120 Si+Si reactions. Specifically, it
happens for the $^{48}$Si+$^{48}$Si reaction (at $E=12.1$ MeV) which
involves
very neutron-rich nuclei. %Expected nuclear physics uncertainties of
%calculated $S(E)$  for such reactions are higher than
%$\widetilde{\eta}_\mathrm{max}+1$  (Sec.\ \ref{s:calculations}) which
%makes our fits quite acceptable.
Although $\widetilde{\eta}_\mathrm{max}+1$ is quite large, our fits are
still acceptable because the expected nuclear physics uncertainties associated
with calculating $S(E)$ for such nuclei are larger
(Sec.\ \ref{s:calculations}). Notice that $\eta_\mathrm{max}$ and
$\widetilde{\eta}_\mathrm{max}$ in Table \ref{tab:fit} can occur for
the same reactions and in the same energy points, and they can even
coincide there. This happens, for instance, for B+N reactions
($\eta_\mathrm{max}=\widetilde{\eta}_\mathrm{max} =0.37$ for
$^{11}$B+$^{13}$N at $E=6.1$ MeV). However, they can also occur for
different reactions and at different $E$ (for instance, in case of
Si+Si we have $\eta_\mathrm{max}=1.12$ for the $^{40}$Si+$^{40}$Si
reaction at $E=14.2$ MeV). We could have changed the fit algorithm
and obtain somewhat smaller $\widetilde{\eta}_\mathrm{max}$ (for
instance, by minimizing $\widetilde{\eta}_\mathrm{rms}$ instead of
$\eta_\mathrm{rms}$) but it would give somewhat larger
${\eta}_\mathrm{rms}$.

%We see that for the Si+Si reactions the fitted values of $S(E)$ do
%not deviate from the calculated ones by more than $\approx$100\%. The
%maximum errors for other reactions are mainly lower, reaching
%$\approx 40$\% for the reactions with light nuclei (e.g., Be+B).
%Root-mean square (rms) relative deviations, given in the ninth column
%of  Table \ref{tab:fit}, are a factor of 3--5 lower than the maximum
%ones. For instance, the rms deviation is 28\% and 8\% for the Si+Si
%and Be+B reactions, respectively. This fit accuracy is quite
%acceptable because it is well within nuclear physics uncertainties of
%the calculated astrophysical $S$-factors (Sec.~\ref{s:calculations}).

\begin{figure}[tbh]
\begin{center}
\includegraphics[width=10.0cm,angle=0]{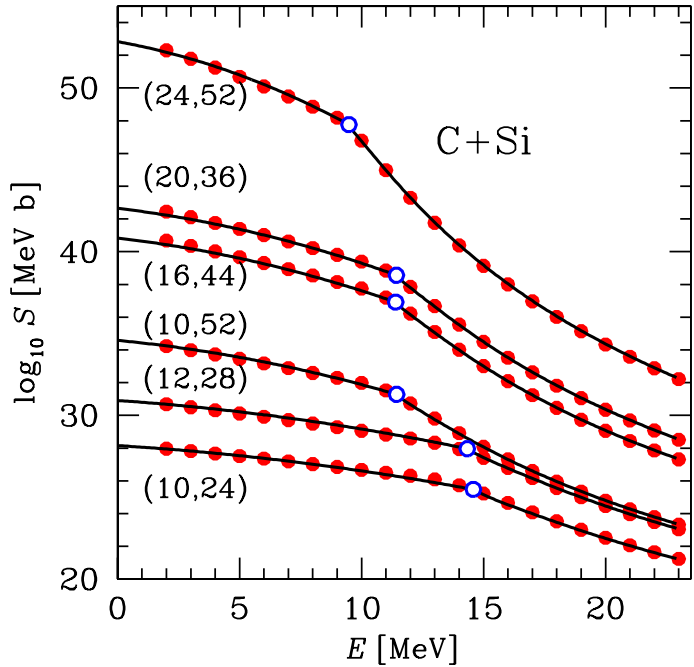}
\caption{(Color online)  $S$-factors for six C+Si reactions $A_1+A_2$
($A_1$ and $A_2$ are given in parentheses). Filled dots are original
calculations (on a rarefied grid of $E$ points); solid lines are our
fits (Table \ref{tab:fit}); open dots show the fit values of $E_C$.}
\label{fig:csi}
\end{center}
\end{figure}

A graphical comparison of the previous fits with calculations for
many reactions is presented in Figs.\ 3 and 4 of Ref.\ \cite{paper2}.
The comparison with the new fits looks the same. For instance, in
Fig.\ \ref{fig:csi} we compare the fitted $S(E)$ (solid lines) with
calculations (dots on a rarefied grid of $E$-points) for six C+Si
reactions. The reactions are labeled by ($A_1,A_2$), where $A_1$ and
$A_2$ are mass numbers of C and Si isotopes, respectively. The lower
line corresponds to the $^{10}$C+$^{24}$Si reaction involving the
lightest (proton rich) nuclei from our collection (Table
\ref{tab:reactions}). The upper line is for the $^{24}$C+$^{52}$Si
reaction involving our most massive (neutron rich) C and Si isotopes.
Other lines refer to intermediate cases, including the
$^{12}$C+$^{28}$Si reaction of most stable nuclei. We see that the
$S$-factors for C+Si reactions vary over many orders of magnitude but
the fit accuracy remains acceptable.

The fit errors for the new $S(E)$-model are generally lower than for
the previous one \cite{paper2}. For instance, for the O+O reactions
we now have the maximum relative fit error
$\eta_\mathrm{max}\approx0.47$ and the rms relative fit error
$\eta_\mathrm{rms}\approx 0.13$ while in Ref.\ \cite{paper2} we had
$\eta_\mathrm{max}\approx 0.50$ and $\eta_\mathrm{rms}\approx 0.14$
(and we now have 7 fit parameters instead of 9). The fit accuracy is
especially improved for reactions involving lower-$Z$ nuclei.

For some reaction types (e.g., for C+C), the fit errors of individual
fits (made separately for every reaction) are noticeably lower than
for all reactions of a given type. This indicates that the fit
formula (\ref{e:ECfit}) for $E_C$ can be further improved. Also, we
expect that the fit quality can be improved by introducing some
slowly varying function $J(E)$ instead of constant $J_0$ at
subbarrier energies in Eq.~(\ref{e:j}), and by elaborating the
description of $S(E)$ at near-barrier energies $|E-E_C|\lesssim
J_0E_0$. However, we think that the accuracy of the new fits is quite
consistent with the quality of the present data (Sec.\
\ref{s:calculations}). The fits give a compact and uniform
description of calculated $S(E)$ for many reactions. They give
reliable $S(E)$ in a wide range of energies $E$ because they are
based on first principles. In addition, they are convenient for
including into computer codes. We warn the readers that the
calculated and fitted $S(E)$ do not take into account resonances.
Therefore, one should add the resonance contribution in modeling
nuclear burning which involves essentially resonant reactions.

Let us remark that our analytic model for $S(E)$ can also be used to
reconstruct the interaction potential $U(r)$ by fitting the $S(E)$
data available from experiment or from calculations. Some examples
have been presented in Ref.\ \cite{paper2}. Another example will be
given below.

%%%%%%%%%%%%%%%%%%%%%%%%%%%%%%%%%%%%%%%%%%%%%%%%%%%%%%%%%%%%%%%%%%
\section{Studying uncertainties of $S(E)$}
\label{s:discussion}
%
%%%%%%%%%%%%%%%%%%%%%%%%%%%%%%%%%%%%%%%%%%%%%%%%%%%%%%%%%%%%%%%%%%

This section illustrates another advantage of our analytic $S(E)$
model -- its ability to study possible uncertainties of astrophysical
reaction rates.

Clearly, any calculation of $S(E)$ contains some uncertainties. First
of all, they can be associated with specific theoretical model. In
our case (Sec.\ \ref{s:calculations}) they are due to using the S\~ao
Paulo potential, the barrier penetration model and the NL3
parametrization for deriving nuclear densities in the RHB theory.
These uncertainties influence an effective potential $U(r)$ and,
hence, $S(E)$. Fitting any given $S(E)$ with our model, one can
estimate the effective potential $U(r)$ [find $\beta$ and $E_C$ in
Eq.~(\ref{e:U})]. Assuming reasonable uncertainties of $\beta$ and
$E_C$ and using our $S(E)$-model again, one can easily estimate the
expected range of $S(E)$ variations.

\begin{figure}[tbh]
\begin{center}
\includegraphics[width=14.0cm,angle=0]{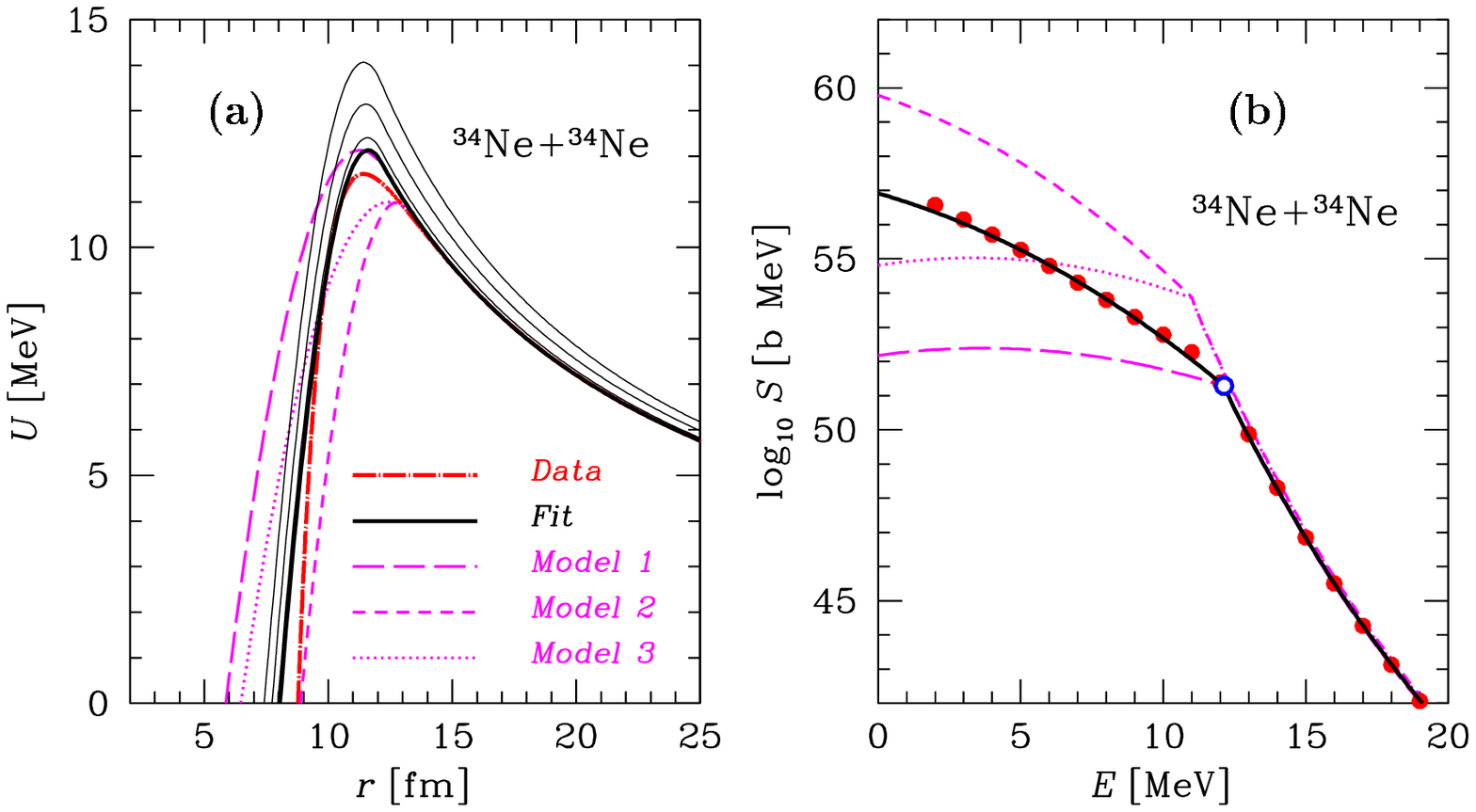}
\caption{(Color online) Effective potentials $U(r)$ (left panel (a))
and respective astrophysical $S(E)$-factors (right panel (b)) for the
$^{34}$Ne+$^{34}$Ne reaction. The dot-dashed line on the left panel
is the original S\~ao Paulo effective potential used in calculations
(it slightly depends on $E$ and is taken at $E=10$ MeV); dots on the
right panel are original calculated $S(E)$ on a rarefied grid of
$E$-points. Thick solid lines show fitted $S(E)$ and $U(r)$
reconstructed from these fits. The three thin solid lines present
appropriate effective potential $V_\mathrm{eff}(r)$ including the
centrifugal term, Eq.\ (\ref{e:Vcentrifug}), for $\ell$=5, 10, and 15
(from bottom to top). Long-dashed, short-dashed, and dotted lines
refer to three models 1, 2, and 3, which illustrate possible effects
of dense matter on the $^{34}$Ne+$^{34}$Ne reaction (see text for
details).}
\label{fig:nene}
\end{center}
\end{figure}

The $S(E)$-model can also be useful to estimate possible effects of
dense matter. Consider, for instance, the heating of the inner crust
of accreting neutron stars in X-ray transients (the so-called deep
crustal heating \cite{hz90,hz03,Brown06}). These transients are
compact binary systems containing a neutron star and a low-mass
companion. The deep crustal heating is thought to occur mainly due to
pycnonuclear reactions in accreted matter when it sinks in the inner
crust under the weight of newly accreted material. The heating can
power \cite{Brown98} thermal surface emission of these neutron stars
that is observed in quiescent states of transients (see, e.g., Refs.\
\cite{Brown06,lh07}). Pycnonuclear reactions occur due to zero-point
vibrations of atomic nuclei in a crystalline lattice. Their physics
is described, for instance, by Salpeter and Van Horn \cite{svh69} and
applied in later work (e.g.\ \cite{gas2005,yak2006,cdi07,cd09} and
references therein). Pycnonuclear reactions occur at high densities
and involve very neutron-rich nuclei \cite{hz90,hz03}, immersed in a
sea of free neutrons available in the inner crust. For example,
consider the powerful $^{34}$Ne+$^{34}$Ne$\to^{68}$Ca reaction at
$\rho \approx 1.7 \times 10^{12}$ g~cm$^{-3}$ in the scenario of
Haensel and Zdunik \cite{hz90}. The fractional number of free
neutrons among all nucleons in the burning layer is 0.39. The
calculations of $S$-factors are performed for fusion of nuclei
unaffected by dense fluid of free neutrons. However, such a fluid can
compress the nuclei, modify their interaction potential $U(r)$, and
hence $S(E)$.

Accurate calculations of $S(E)$ under these conditions have not been
performed; they are complicated and do deserve a special study.
Nevertheless, the $S(E)$-model allows us to estimate the range of
expected $S(E)$ changes. This is illustrated in Fig.\ \ref{fig:nene}
which shows the model effective potentials (left panel (a)) and
respective astrophysical factors $S(E)$ for the $^{34}$Ne+$^{34}$Ne
reaction (right panel (b)) without and with possible effects of dense
matter.

The filled dots in the right panel of Fig.\ \ref{fig:nene} are our
calculated $S(E)$ on a rarefied grid of energies $E$ and the thick
solid line is our fit (just like in Fig.\ \ref{fig:csi} for some C+Si
reactions). The dot-dashed line in the left panel is the effective
potential $U(r)$ used in calculations. According to the S\~ao Paulo
model, the theoretical effective potential slightly depends on $E$.
In Fig.\ \ref{fig:nene} it is taken at $E=10$ MeV. The thick solid
line is our model effective potential $U(r)$, which is given by
Eq.~(\ref{e:U}) and plotted for the parameters $E_C=12.137$ MeV and
$\delta=0.0441$ inferred from the fit (Table \ref{tab:fit}). We see
that fitting the available (here -- calculated) $S(E)$ data with our
analytic model allows us to reconstruct $U(r)$ with sufficiently good
precision. More examples of successful reconstructions are given in
Ref.\ \cite{paper2}. The three thin solid lines in Fig.\ \ref{fig:nene}
illustrate our discussion on the contribution of $\ell>0$ waves to nuclear
fusion (Sec.\ \ref{s:ell}). These solid lines represent the reconstructed
effective potential, $V_\mathrm{eff}(r)$, including the centrifugal terms
$\ell=$5, 10, 15 (bottom to top). A few lowest $\ell$-waves
penetrate the barrier almost with the same efficiency as the $s$-wave
and contribute to $S(E)$ even at low $E$.

Other curves in Fig.\ \ref{fig:nene} demonstrate possible effects of
dense matter. Nuclear reaction rates are proportional to the value of
$S(E)$ at a typical reaction energy $E$. In pycnonuclear reactions
the energy $E$ is
%taken from
%zero-point vibrations, $E \sim \hbar \omega_p$
rather low \cite{svh69},
%, $\omega_p$ being the ion plasma frequency.
%For the $^{34}$Ne+$^{34}$Ne reaction at $\rho \approx 1.7 \times
%10^{12}$ g~cm$^{-3}$, we have $\hbar \omega_p \sim 0.2$ MeV
so that the reaction rate is actually determined by $S(0)$. It is
reasonable to expect that the presence of free neutrons between the
reacting nuclei broadens and/or lowers the maximum of $U(r)$. In
Fig.\ \ref{fig:nene} we consider three models of this phenomenon
labeled as 1, 2, and 3. Model 1 (long-dashed lines) assumes extra
broadening of the $U(r)$ peak at the same height ($E_C=12.137$ MeV as
before but larger $\delta=0.1$). The barrier $U(r)$ becomes thicker
(left panel) which lowers $S(0)$ by about eight orders of magnitude.
The reaction rate will be strongly suppressed which may cause
\cite{ygw06} a delayed $^{34}$Ne burning after accretion stops in an
X-ray transient. Model 2 (short dashed lines) assumes the same
curvature of the $U(r)$ peak ($\delta=0.0441$) as for the initial
potential but lower maximum, $E_C=11$ MeV. The lower barrier is
naturally more transparent, which increases the initial value of
$S(0)$ by about three orders of magnitude; the $^{34}$Ne burning will
react quicker to variations of accretion rate. This possibility has
also been considered in Ref.\ \cite{ygw06}. Finally, model 3 (dotted
lines) assumes that the medium effects simultaneously lower and
broaden the barrier ($E_C=11$ MeV, $\delta=0.1$). The lowering makes
the barrier more transparent while the broadening makes it more
opaque. In this example, the broadening wins so that $S(0)$ is about
two orders of magnitude smaller than the initial value. Therefore, we
may expect that the medium effects can greatly enhance or suppress
pycnonuclear reaction rates and the net effect is not clear. Also,
the presence of free neutrons between the reacting nuclei may change
$U(r)$ in such a way that the approximation (\ref{e:U}) becomes poor.
In addition, the theoretical expression for the pycnonuclear reaction
rate through $S(E)$ contains serious uncertainties
\cite{yak2006,ygw06} which complicate the problem. Further studies
are required to clarify these points.

Let us add that nuclear reaction rates in dense stellar matter
(especially, in the cores of white dwarfs and envelopes of neutron
stars \cite{st83}) can be greatly affected not only by the transition
to pycnonuclear burning regime but also by plasma screening of the
Coulomb interaction. The plasma effects were described by Salpeter
and Van Horn \cite{svh69} (also see
\cite{gas2005,yak2006,cdi07,cd09,pc12} and references therein). They
modify the interaction potential $U(r)$ but mainly at sufficiently
large $r$, typically larger than nuclear scales, whereas we discuss
the nuclear physics effects which influence $U(r)$ at smaller $r$. It
is commonly thought that the plasma physics and nuclear physics
effects are distinctly different and can be considered separately.
However, as we noticed in \cite{paper2}, in dense and not very hot
stellar matter both effects become interrelated and should be studied
together.

%%%%%%%%%%%%%%%%%%%%%%%%%%%%%%%%%%%%%%%%%%%%%%%
\section{Conclusions}
\label{s:concl}
%%%%%%%%%%%%%%%%%%%%%%%%%%%%%%%%%%%%%%%%%%%%%%%

We have performed new calculations  and created a large database of
the $S$-factors for about 5,000 fusion reactions (Sec.\
\ref{s:calculations}) involving various isotopes of Be, B, C, N, O,
F, Ne, Na, Mg, and Si located between proton and neutron drip lines.
Note that the drip lines are obtained in the RHB calculations with
the NL3 parametrization. The $S$-factors were calculated using the
S\~ao Paulo method and the barrier penetration model with nuclear
densities obtained in the RHB calculations. We have elaborated
 and simplified (Sec.~\ref{s:model}) our model \cite{paper2} for
describing the astrophysical $S$-factor as a function of
center-of-mass energy $E$ of reacting nuclei for non-resonant fusion
reactions. Our main results are:

\begin{itemize}

\item For any reaction, we present $S(E)$ in a simple analytic form in
terms of three parameters (Sec.\ \ref{s:instruction}). They are
$E_C$, the height of the Coulomb barrier; $\delta$ that describes the
broadening of the peak of the effective barrier potential $U(r)$; and
$j_0$ that measures the contribution of $\ell>0$ waves at subbarrier
energies. Analytic fits are expected to be sufficiently accurate for
energies below and above $E_C$ (up to a few $E_C$).

\item We succeeded to fit all our $S(E)$-data with only 7 fit parameters
for any group of reactions involving isotopes of the same elements
(Sec.~\ref{s:Fits}, Table \ref{tab:fit}). The fit accuracy is well
within estimated nuclear-physics uncertainties of calculated $S(E)$
(Sec.~\ref{s:calculations}).

\item In this way we obtain a simple, accurate, uniform and ultracompact
database for calculating $S(E)$; the instructions for users are given
in Sec.\ \ref{s:instruction}. It is easy to implement the database
into computer codes (especially in network-type ones) which calculate
nuclear reaction rates and simulate various nuclear burning phenomena
in astrophysical environment.

\item In comparison with our previous $S(E)$ model \cite{paper2}, the
present version is simpler, more accurate and reliable. We have
simplified the analytic expression for $S(E)$ at subbarrier energies
(Sec.\ \ref{s:lowE}). We have clarified the contribution of $\ell>0$
waves to $S(E)$ and introduced the parameter ($J_0$ or $j_0$) that
accounts for this contribution (Sec.\ \ref{s:ell}). We have replaced
a phenomenological analytic expression for $S(E)$ at energies above
the barrier by a rigorous semi-classical expression (Sec.\
\ref{s:highE}). These modifications have allowed us to reduce the number
of fit parameters (now 3 parameters instead of 4 for any given
$S(E)$, and 7 parameters instead of 9 for any group of reactions
involving isotopes of the same elements).

\item We have discussed (Sec.\ \ref{s:discussion}) the possibility
of using our model for estimating uncertainties of $S(E)$ values and
for studying the effects of dense matter on $S(E)$. For illustration,
we have analyzed the range of variations of $S(E)$ due to in-medium
deformations of interaction potential $U(r)$ for the pycnonuclear
$^{34}$Ne+$^{34}$Ne reaction in the inner crust of an accreting
neutron star (with the conclusion that the variations can reach
several orders of magnitude).

\end{itemize}

As detailed in Ref.\ \cite{paper2}, the analytic $S(E)$ model is
practical for describing large uniform sets of $S(E)$ data. The
parameters of the model can be interpolated from one reaction to
another which can be useful in the case when some $S(E)$ data are not
available. The functional form of our analytical $S(E)$ is flexible
enough to describe different behaviors of $S(E)$ at low energies
\cite{paper2}. Fitting a given $S(E)$ (computed or measured in
laboratory) with our analytic model can be used to reconstruct the
effective potential $U(r)$ (Sec.\ \ref{s:discussion}; also see Ref.\
\cite{paper2}). There is no doubt that the present model can be
improved (Sec.\ \ref{s:instruction}; Ref.\ \cite{paper2}),
particularly, by complicating the interaction potential $U(r)$.
However, this would require the introduction of new parameters which
would complicate the model. This will improve the description of
$S(E)$ in some particular cases but the most attractive features of
the model -- its simplicity and universality -- would be lost.

\begin{acknowledgments}
This work is partially supported by JINA (PHYS-0822648) and by the
U.S. Department of Energy under the grant DE-FG02-07ER41459
(Mississippi State University). AIC and DGY acknowledge support from
RFBR (grant 11-02-00253-a) and from the State Program of ``Leading
Scientific Schools of Russian Federation'' (grant NSh 4035.2012.2).
AIC acknowledges support of the Dynasty Foundation, of the President
grant for young Russian scientists (MK-857.2012.2), and of the RAS
Presidium Program ``Support for Young Scientists''; DGY acknowledges
support of RFBR (grant 11-02-12082-ofi-m-2011) and Ministry of
Education and Science of Russian Federation (contract
11.G34.31.0001).
\end{acknowledgments}

\begin{center}
\begin{table}
\caption{Fusion reactions $(A_1,Z_1)+(A_2,Z_2)$ under consideration.
See text for details} \label{tab:reactions}
\begin{tabular}{c c c c c}
\hline \hline Reaction & $\quad A_1 \quad $ & $\quad A_2 \quad $ &
~$E_\mathrm{max}$~ & Number  \\
type   &       &        &  MeV & of cases  \\
\hline \hline
%\endfirsthead
%
%\caption{Fusion reactions $(A_1,Z_1)+(A_2,Z_2)$ under consideration
%(continued)}\\
%\hline \hline Reaction & $\quad A_1 \quad $ & $\quad A_2 \quad $ &
%~$E_\mathrm{max}$~ & Number  \\
%type   &       &        &  MeV & of cases  \\
%\hline \hline \endhead
%
Be+Be &  8--14 &  8--14 & 15.9 & 10  \\
Be+B  &  8--14 &  9--21 & 16.9 & 28  \\
Be+C  &  8--14 & 10--24 & 16.9 & 32  \\
Be+N  &  8--14 & 11--27 & 17.9 & 36  \\
Be+O  &  8--14 & 12--28 & 18.9 & 36  \\
Be+F  &  8--14 & 17--29 & 18.9 & 28  \\
Be+Ne &  8--14 & 18--40 & 19.9 & 48  \\
Be+Na &  8--14 & 19--43 & 21.9 & 52  \\
Be+Mg &  8--14 & 20--46 & 22.9 & 56  \\
Be+Si &  8--14 & 24--52 & 23.9 & 60  \\
B+B   &  9--21 &  9--21 & 15.9 & 28  \\
B+C   &  9--21 & 10--24 & 16.8 & 56  \\
B+N   &  9--21 & 11--27 & 17.8 & 63  \\
B+O   &  9--21 & 12--28 & 18.8 & 63  \\
B+F   &  9--21 & 17--29 & 18.8 & 49  \\
B+Ne  &  9--21 & 18--40 & 19.8 & 84  \\
B+Na  &  9--21 & 19--43 & 21.8 & 91  \\
B+Mg  &  9--21 & 20--46 & 22.8 & 98  \\
B+Si  &  9--21 & 24--52 & 23.8 & 105 \\
C+C   & 10--24 & 10--24 & 17.9 & 36  \\
C+N   & 10--24 & 11--27 & 19.8 & 72  \\
C+F   & 10--24 & 17--29 & 20.8 & 56  \\
C+O   & 10--24 & 12--28 & 17.9 & 72  \\
C+Ne  & 10--24 & 18--40 & 19.9 & 96  \\
C+Na  & 10--24 & 19--43 & 21.8 & 104 \\
C+Mg  & 10--24 & 20--46 & 19.9 & 112 \\
C+Si  & 10--24 & 24--52 & 24.8 & 120 \\
N+N   & 11--27 & 11--27 & 17.8 & 45  \\
N+O   & 11--27 & 12--28 & 19.8 & 81  \\
N+F   & 11--27 & 17--29 & 20.8 & 63  \\
N+Ne  & 11--27 & 18--40 & 21.8 & 108 \\
N+Na  & 11--27 & 19--43 & 21.9 & 117 \\
N+Mg  & 11--27 & 20--46 & 21.9 & 126 \\
N+Si  & 11--27 & 24--52 & 24.8 & 135 \\
O+O   & 12--28 & 12--28 & 19.9 & 45  \\
O+F   & 12--28 & 17--29 & 21.8 & 63  \\
O+Ne  & 12--28 & 18--40 & 21.9 & 108  \\
O+Na  & 12--28 & 19--43 & 21.8 & 117  \\
O+Mg  & 12--28 & 18--46 & 21.9 & 126  \\
O+Si  & 12--28 & 24--52 & 24.8 & 135 \\
F+F   & 17--29 & 17--29 & 19.8 & 28  \\
F+Ne  & 17--29 & 18--40 & 21.9 & 84  \\
F+Na  & 17--29 & 19--43 & 24.9 & 91  \\
F+Mg  & 17--29 & 20--46 & 24.9 & 98  \\
F+Si  & 17--29 & 24--52 & 29.9 & 105 \\
Ne+Ne & 18--40 & 18--40 & 21.9 & 78  \\
Ne+Na & 18--40 & 19--43 & 29.9 & 156 \\
Ne+Mg & 18--40 & 20--46 & 24.9 & 168  \\
Ne+Si & 18--40 & 24--52 & 29.8 & 180 \\
Na+Na & 19--43 & 19--43 & 21.8 & 91  \\
Na+Mg & 19--43 & 20--46 & 29.9 & 182 \\
Na+Si & 19--43 & 24--52 & 37.9 & 195 \\
Mg+Mg & 20--46 & 20--46 & 29.9 & 105  \\
Mg+Si & 20--46 & 24--52 & 39.8 & 210 \\
Si+Si & 24--52 & 24--52 & 39.8 & 120 \\
\hline \hline
\end{tabular}
\end{table}
\end{center}

%\end{document}
%\input{taba1.tex}

\begin{center}
\begin{table}
\caption{Fit parameters of $S(E)$ for reactions $(A_1,Z_1)+(A_2,Z_2)$
} \label{tab:fit}
\begin{tabular}{c r r r r r r r c c c }
 \hline \hline Reaction & ~~~$R$~~~~ & ~~$\Delta
R_{1a}$~~~ & ~~$\Delta R_{2a}$~~~& ~~$\Delta R_{1b}$~~~ & ~~$\Delta
R_{2b}$~~~ & ~~~~~$\delta$~~~~~& ~~~~$j_0$~~~~~&
$\eta_\mathrm{rms}$~~~ &
$\eta_\mathrm{max}$~~~& $\widetilde{\eta}_\mathrm{max}$ \\
type   & fm~~~ & fm~~~  & fm~~~ & fm~~~  & fm~~~ &
 &
 & ~~~~   &
~~ &  \\
%\hline \hline \endfirsthead \caption{Fit parameters of $S(E)$ for
%reactions $(A_1,Z_1)+(A_2,Z_2)$ (continued)
%}  \\
%\hline \hline Reaction & ~~~$R$~~~~ & ~~$\Delta R_{1a}$~~~ &
%~~$\Delta R_{2a}$~~~& ~~$\Delta R_{1b}$~~~ & ~~$\Delta R_{2b}$~~~ &
%~~~~~$\delta$~~~~~& ~~~~$j_0$~~~~~& $\eta_\mathrm{rms}$~~~ &
%$\eta_\mathrm{max}$~~~& $\widetilde{\eta}_\mathrm{max}$ \\
%type   & fm~~~ & fm~~~  & fm~~~ & fm~~~  & fm~~~ &
%   &
%   &
%~~~~   &
%~~ &  \\
%\hline \hline \endhead
%
\hline
Be+Be &
  7.5010 &
  0.2480 &
  0.2480 &
  0.1557 &
  0.1557 &
  0.0330 &
  0.7453 &
   0.08 &
   0.37 &
   0.59 \\
Be+B &
  7.5065 &
  0.2547 &
  0.2223 &
 --1.7469&
  0.0635 &
  0.0370 &
  0.7814 &
   0.08 &
   0.40 &
   0.54 \\
Be+C &
  7.6982 &
  0.2543 &
  0.1877 &
  2.0844 &
 --0.0012 &
  0.0441 &
  0.7349 &
   0.09 &
   0.47 &
   0.67 \\
Be+N &
  7.9324 &
  0.2523 &
  0.1546 &
  7.7591 &
 --0.0233 &
  0.0484 &
  0.7446 &
   0.10 &
   0.41 &
   0.70 \\
Be+O &
  8.0708 &
  0.2507 &
  0.1346 &
 --0.0738 &
 --0.0271 &
  0.0509 &
  0.7699 &
   0.11 &
   0.44 &
   0.80 \\
Be+F &
  8.1585 &
  0.2510 &
  0.1201 &
 --0.6214 &
 --0.0272 &
  0.0509 &
  0.8117 &
   0.12 &
   0.48 &
   0.82 \\
Be+Ne &
  8.1485 &
  0.2510 &
  0.1270 &
 --1.5671 &
  0.0007 &
  0.0513 &
  0.8697 &
   0.13 &
   0.59 &
   1.26 \\
Be+Na &
  8.2139 &
  0.2494 &
  0.1161 &
 --1.3477 &
 --0.0059 &
  0.0505 &
  0.9065 &
   0.13 &
   0.52 &
   1.10 \\
Be+Mg &
  8.2734 &
  0.2481 &
  0.1067 &
 --0.5794 &
 --0.0128 &
  0.0499 &
  0.9528 &
   0.14 &
   0.57 &
   1.30 \\
Be+Si &
  8.3390 &
  0.2462 &
  0.0937 &
  1.8651 &
 --0.0126 &
  0.0481 &
  1.0538 &
   0.16 &
   0.62 &
   1.26 \\
%%%%%%%%%%%%%%%%%%%%%%%%%%%%%%%%%%%%%%%%%%
B+B &
  7.6600 &
  0.2175 &
  0.2175 &
  0.0511 &
  0.0511 &
  0.0459 &
  0.7772 &
   0.07 &
   0.51 &
   0.51 \\
B+C &
  7.8155 &
  0.2155 &
  0.1839 &
  0.0414 &
 --0.0148 &
  0.0511 &
  0.7919 &
   0.07 &
   0.40 &
   0.43 \\
B+N &
  8.0004 &
  0.2132 &
  0.1522 &
  0.0334 &
 --0.0359 &
  0.0523 &
  0.8523 &
   0.07 &
   0.37 &
   0.37 \\
B+O &
  8.1037 &
  0.2117 &
  0.1331 &
  0.0338 &
 --0.0378 &
  0.0520 &
  0.9001 &
   0.08 &
   0.47 &
   0.47 \\
B+F &
  8.1700 &
  0.2114 &
  0.1188 &
  0.0349 &
 --0.0379 &
  0.0502 &
  0.9575 &
   0.09 &
   0.48 &
   0.51 \\
B+Ne &
  8.1755 &
  0.2088 &
  0.1244 &
  0.0113 &
 --0.0098 &
  0.0501 &
  1.0353 &
   0.12 &
   0.57 &
   1.08 \\
B+Na &
  8.2418 &
  0.2075 &
  0.1137 &
  0.0082 &
 --0.0160 &
  0.0492 &
  1.0885 &
   0.11 &
   0.56 &
   0.85 \\
B+Mg &
  8.3033 &
  0.2064 &
  0.1047 &
  0.0061 &
 --0.0211 &
  0.0485 &
  1.1431 &
   0.12 &
   0.58 &
   1.03 \\
B+Si &
  8.3977 &
  0.2048 &
  0.0921 &
  0.0035 &
 --0.0201 &
  0.0472 &
  1.2073 &
   0.14 &
   0.59 &
   1.05 \\
%%%%%%%%%%%%%%%%%%%%%%%%%%%%%%%%%%%%%%%%%%
C+C &
  7.8843 &
  0.1836 &
  0.1836 &
 --0.0107 &
 --0.0107 &
  0.0524 &
  0.8476 &
   0.08 &
   0.49 &
   0.49 \\
C+N &
  8.0464 &
  0.1816 &
  0.1516 &
 --0.0181 &
 --0.0375 &
  0.0515 &
  0.9341 &
   0.08 &
   0.45 &
   0.45 \\
C+O &
  8.1523 &
  0.1806 &
  0.1324 &
 --0.0173 &
 --0.0393 &
  0.0507 &
  0.9647 &
   0.10 &
   0.51 &
   0.59 \\
C+F &
  8.2103 &
  0.1804 &
  0.1184 &
 --0.0164 &
 --0.0383 &
  0.0487 &
  1.0425 &
   0.10 &
   0.53 &
   0.58 \\
C+Ne &
  8.2146 &
  0.1790 &
  0.1239 &
 --0.0293 &
 --0.0089 &
  0.0487 &
  1.1230 &
   0.15 &
   0.64 &
   1.34 \\
C+Na &
  8.2839 &
  0.1780 &
  0.1132 &
 --0.0306 &
 --0.0165 &
  0.0479 &
  1.1797 &
   0.14 &
   0.63 &
   1.01 \\
C+Mg &
  8.3785 &
  0.1772 &
  0.1043 &
 --0.0305 &
 --0.0214 &
  0.0477 &
  1.1405 &
   0.16 &
   0.61 &
   1.24 \\
C+Si &
  8.4392 &
  0.1763 &
  0.0916 &
 --0.0320 &
 --0.0205 &
  0.0461 &
  1.3190 &
   0.16 &
   0.66 &
   1.28 \\
%%%%%%%%%%%%%%%%%%%%%%%%%%%%%%%%%%%%%%%%%%
N+N &
  8.2069 &
  0.1504 &
  0.1504 &
 --0.0415 &
 --0.0415 &
  0.0503 &
  1.0043 &
   0.08 &
   0.37 &
   0.44 \\
N+O &
  8.2988 &
  0.1494 &
  0.1316 &
 --0.0425 &
 --0.0425 &
  0.0492 &
  1.0732 &
   0.10 &
   0.45 &
   0.62 \\
N+F &
  8.3606 &
  0.1493 &
  0.1176 &
 --0.0423 &
 --0.0403 &
  0.0473 &
  1.1550 &
   0.09 &
   0.44 &
   0.53 \\
N+Ne &
  8.3526 &
  0.1485 &
  0.1227 &
 --0.0539 &
 --0.0112 &
  0.0472 &
  1.2964 &
   0.15 &
   0.58 &
   1.31 \\
N+Na &
  8.4329 &
  0.1477 &
  0.1121 &
 --0.0559 &
 --0.0190 &
  0.0466 &
  1.3367 &
   0.14 &
   0.53 &
   0.97 \\
N+Mg &
  8.5052 &
  0.1471 &
  0.1032 &
 --0.0572 &
 --0.0235 &
  0.0462 &
  1.3764 &
   0.18 &
   0.53 &
   1.14 \\
N+Si &
  8.5820 &
  0.1463 &
  0.0907 &
 --0.0595 &
 --0.0222 &
  0.0448 &
  1.5534 &
   0.16 &
   0.63 &
   1.08 \\
%%%%%%%%%%%%%%%%%%%%%%%%%%%%%%%%%%%%%%%%%%
O+O &
  8.3972 &
  0.1309 &
  0.1309 &
 --0.0439 &
 --0.0439 &
  0.0483 &
  1.1199 &
   0.13 &
   0.47 &
   0.87 \\
O+F &
  8.4602 &
  0.1306 &
  0.1169 &
 --0.0430 &
 --0.0395 &
  0.0464 &
  1.2118 &
   0.12 &
   0.46 &
   0.81 \\
O+Ne &
  8.4521 &
  0.1304 &
  0.1219 &
 --0.0524 &
 --0.0107 &
  0.0464 &
  1.3703 &
   0.18 &
   0.64 &
   1.67 \\
O+Na &
  8.5366 &
  0.1297 &
  0.1113 &
 --0.0540 &
 --0.0192 &
  0.0460 &
  1.4007 &
   0.17 &
   0.59 &
   1.42 \\
O+Mg &
  8.5962 &
  0.1292 &
  0.1025 &
 --0.0553 &
 --0.0238 &
  0.0453 &
  1.4973 &
   0.18 &
   0.63 &
   1.68 \\
O+Si &
  8.6699 &
  0.1286 &
  0.0900 &
 --0.0573 &
 --0.0224 &
  0.0440 &
  1.7240 &
   0.20 &
   0.67 &
   1.61 \\
%%%%%%%%%%%%%%%%%%%%%%%%%%%%%%%%%%%%%%%%%%
F+F &
  8.5607 &
  0.1169 &
  0.1169 &
 --0.0355 &
 --0.0355 &
  0.0452 &
  1.1785 &
   0.12 &
   0.46 &
   0.69 \\
F+Ne &
  8.5166 &
  0.1169 &
  0.1221 &
 --0.0471 &
 --0.0077 &
  0.0449 &
  1.4636 &
   0.19 &
   0.66 &
   1.85 \\
F+Na &
  8.5808 &
  0.1164 &
  0.1113 &
 --0.0484 &
 --0.0184 &
  0.0441 &
  1.6063 &
   0.16 &
   0.63 &
   1.25 \\
F+Mg &
  8.6505 &
  0.1159 &
  0.1024 &
 --0.0489 &
 --0.0232 &
  0.0437 &
  1.6786 &
   0.18 &
   0.61 &
   1.53 \\
F+Si &
  8.7286 &
  0.1154 &
  0.0899 &
 --0.0497 &
 --0.0220 &
  0.0424 &
  1.9341 &
   0.18 &
   0.71 &
   1.43 \\
%%%%%%%%%%%%%%%%%%%%%%%%%%%%%%%%%%%%%%%%%%
Ne+Ne &
  8.4649 &
  0.1214 &
  0.1214 &
 --0.0145 &
 --0.0145 &
  0.0441 &
  1.8709 &
   0.25 &
   0.91 &
   4.00 \\
Ne+Na &
  8.5397 &
  0.1205 &
  0.1106 &
 --0.0185 &
 --0.0263 &
  0.0435 &
  2.0638 &
   0.21 &
   0.82 &
   2.95 \\
Ne+Mg &
  8.6277 &
  0.1200 &
  0.1020 &
 --0.0189 &
 --0.0297 &
  0.0433 &
  2.0390 &
   0.27 &
   0.80 &
   3.53 \\
Ne+Si &
  8.7139 &
  0.1192 &
  0.0896 &
 --0.0193 &
 --0.0273 &
  0.0423 &
  2.3281 &
   0.26 &
   0.99 &
   3.17 \\
%%%%%%%%%%%%%%%%%%%%%%%%%%%%%%%%%%%%%%%%%%
Na+Na &
  8.6464 &
  0.1100 &
  0.1100 &
 --0.0265 &
 --0.0265 &
  0.0434 &
  1.9895 &
   0.25 &
   0.77 &
   2.52 \\
Na+Mg &
  8.7081 &
  0.1093 &
  0.1013 &
 --0.0285 &
 --0.0309 &
  0.0429 &
  2.1976 &
   0.22 &
   0.79 &
   2.52 \\
Na+Si &
  8.7972 &
  0.1086 &
  0.0890 &
 --0.0296 &
 --0.0284 &
  0.0419 &
  2.4993 &
   0.23 &
   0.88 &
   2.68 \\
%%%%%%%%%%%%%%%%%%%%%%%%%%%%%%%%%%%%%%%%%%
Mg+Mg &
  8.7791 &
  0.1009 &
  0.1009 &
 --0.0311 &
 --0.0311 &
  0.0425 &
  2.2903 &
   0.24 &
   0.80 &
   3.01 \\
Mg+Si &
  8.8704 &
  0.1002 &
  0.0886 &
 --0.0326 &
 --0.0290 &
  0.0416 &
  2.6253 &
   0.24 &
   0.90 &
   3.08 \\
%%%%%%%%%%%%%%%%%%%%%%%%%%%%%%%%%%%%%%%%%%
Si+Si &
  8.9765 &
  0.0880 &
  0.0880 &
 --0.0292 &
 --0.0292 &
  0.0409 &
  2.8162 &
   0.28 &
   1.12 &
   4.40 \\
%%%%%%%%%%%%%%%%%%%%%%%%%%%%%%%%%%%%%%%%%%
  \hline \hline
\end{tabular}
\end{table}
\end{center}

%\newpage

%\input{taba2.tex}


\begin{thebibliography}{222}
%

%1
\bibitem{bbfh57}
E.~M.\ Burbidge, G.~R.\ Burbidge, W.~A.\ Fowler, and F.~Hoyle, Rev.\
Mod.\ Phys.\ {\bf 29}, 547 (1957).

%2
\bibitem{fh64}
W.~A.~Fowler and F.~Hoyle, Astrophys.\ J.\ Suppl.\ {\bf 9}, 201
(1964); Appendix C.

%3
\bibitem{clayton83}
D.~D.\ Clayton, {\em Principles of Stellar Evolution and
Nucleosynthesis} (University of Chicago Press, Chicago, 1983).

%4
\bibitem{st83}
S.\ L.\ Shapiro and S.\ A.\ Teukolsky, {\em Black Holes, White
Dwarfs, and Neutron Stars} (Wiley-Interscience, New York, 1983).

\bibitem{svh69}
E.~E.~Salpeter and H.~M.~Van Horn, Astrophys.\ J.\ {\bf 155}, 183
(1969).

\bibitem{yak2006}
D.~G. Yakovlev, L.~R. Gasques, M. Beard, M. Wiescher, and A.~V.
Afanasjev, Phys. Rev. C {\bf 74}, 035803 (2006).

\bibitem{cd09}
A.~I. Chugunov and H.~E. DeWitt,
%Coulomb tunneling for fusion reactions in dense matter:
%Path integral Monte Carlo versus mean field,
Phys. Rev. C {\bf 80}, 014611 (2009).

%5
\bibitem{NiWo97}
J.~C.~Niemeyer and S.~E.~Woosley, Astrophys.\ J.\ {\bf 475} 740
(1997).

%6
\bibitem{hoeflich06}
P.~H\"oflich,
% Physics of type Ia supernovae,
Nucl.\ Phys.\ A {\bf 777}, 579 (2006).

%7
\bibitem{vankerkwijketal2010}
M.~H.\ van Kerkwijk, P.\ Chang, S.\ Justham,
%Sub-Chandrasekhar White Dwarf Mergers as the Progenitors of Type Ia
%Supernovae
Astrophys.\ J.\ {\bf 722},  L157 (2010).

%8
\bibitem{sb06}
T.~Strohmayer and L. Bildsten, in {\em Compact Stellar X-Ray
Sources}, eds.\ W.~H.~G. Lewin, M.\ Van der Klis (Cambridge
University Press, Cambridge, 2006), p.\ 113.

%9
\bibitem{schatz03}
H. Schatz , L. Bildsten, and A. Cumming, Astrophys. J. {\bf 583}, L87
(2003).

%10
\bibitem{cummingetal05}
A.~Cumming,  J.\ Macbeth, J.~J.~M.\ in 't Zand, and D.\ Page,
Astrophys.\ J.\ {\bf 646}, 429 (2006).

%11
\bibitem{Brown06}
S.\ Gupta, E.~F.~Brown, H.\ Schatz, P.\ M\"oller, and K.-L.\ Kratz,
Astrophys.\ J.\ {\bf 662}, 1188 (2007).

%12
\bibitem{cooperetal09}
R.~L.\ Cooper, A.~W.\ Steiner, and E.~F.\ Brown,
%Possible Resonances
%in the 12C + 12C Fusion Rate and Superburst Ignition.
Astrophys.\ J.\ {\bf 702}, 660 (2009).


\bibitem{hz90}
P.\ Haensel and J.~L.\ Zdunik, Astron.\ Astrophys.\ {\bf 229}, 117
(1990).

\bibitem{hz03}
P.\ Haensel and J.~L.\ Zdunik, Astron.\ Astrophys.\ {\bf 404}, L33
(2003).

\bibitem{Brown98}
E.~F.~Brown and L.~Bildsten, Astrophys.\ J.\ {\bf 496}, 915 (1998).

%13
\bibitem{pgw06}
D.~Page, U.~Geppert, and F.\ Weber, Nucl.\ Phys. A,\ {\bf 777}, 497
(2006).


\bibitem{lh07}
K.~P.\ Levenfish and P.~Haensel, Astrophys.\ Space Sci.\ {\bf 308},
457 (2007).

\bibitem{bk09}
E.~F.\ Brown and A.\ Cumming,
%Mapping Crustal Heating with the
%Cooling Light Curves of Quasi-Persistent Transients
Astrophys.\ J.\ {\bf 698}, 1020 (2009).

\bibitem{MW-08} M.\ Beard et al.,
%, M.\ Wiescher, L.\ Gasques, A.\ Afanasjev,
%D.\ Yakovlev, E.\ Brown, K.~Y.\ Lau, H.\ Schatz, W.\ R.\ Hix, S.\ Gupta, P.\ M{\"o}ller, A.\
%Steiner, and K.-L.\ Kratz,
Proc.\ of Science (Nuclei in the Cosmos X)  {\bf 182}, 1 (2008);
available in electronic form at
[pos.sissa.it/archive/conferences/053/182/NIC\%20X\_182.pdf].

\bibitem{paper1}
M. Beard, A.~V. Afanasjev, L.~C. Chamon, L.~R. Gasques, M. Wiescher,
and D. G. Yakovlev, Atomic Data Nucl.\ Data Tables {\bf 96}, 541 (2010). %[arXiv:1002.0741].

\bibitem{paper2}
D.~G.\ Yakovlev, M.\ Beard, L.~R.\ Gasques, and M.~Wiescher,
%Simple analytic model for astrophysical S factors,
Phys.\ Rev.\ C {\bf 82}, 044609 (2010).

%\bibitem{LLQM}
%L.~D.\ Landau and E.~M.~Lifshitz, {\em Quantum Mechanics} (Pergamon,
%Oxford, 1976).

\bibitem{gas2005}
L.~R. Gasques, A.~V. Afanasjev, E.~F. Aguilera, M. Beard, L.~C.
Chamon, P. Ring, M. Wiescher, and D.~G. Yakovlev, Phys. Rev. C {\bf
72}, 025806 (2005).

\bibitem{VALR.05}
D.\ Vretenar, A.~V.\ Afanasjev, G.~A.\ Lalazissis, and P.\ Ring,
Phys.\ Rep.\ {\bf 409},  101  (2005).

\bibitem{NL3}
G.~A.\ Lalazissis, J.\ K\"{o}nig, and P.\ Ring, Phys.\ Rev.\ {\bf
C55}, 540 (1997).


\bibitem{saoPauloTool}
L.~R. Gasques, A.~V. Afanasjev, M. Beard, J. Lubian, T. Neff, M.
Wiescher, and D.~G. Yakovlev, Phys. Rev. C {\bf 76}, 045802 (2007).

%\bibitem{Jiang07}
%C.~L. Jiang, K.~E. Rehm, B.~B. Back, and R.~V.~F. Janssens, Phys.
%Rev. C {\bf 75}, 015803 (2007).

%\bibitem{hindrance}
%L.~R.~Gasques, E.~F. Brown, A.~Chieffi, C.~L. Jiang, M.~Limongi,
%C.~Rolfs, M.~Wiescher, and D.~G. Yakovlev, Phys. Rev. C {\bf 76},
%035802 (2007).

\bibitem{vas81}
L.~C.\ Vaz, J.~M.\ Alexander, and G.~R.\ Satchler,
% Fusion barriers, empirical and theoretical:
% evidence for dynamic deformation in subbarier fusion
Phys.\ Rep.\ {\bf 69}, 373 (1981).

\bibitem{leandro04}
L.~R.\ Gasques, L.~C.\ Chamon, D.\ Pereira, M.~A.~G.\ Alvarez, E.~S.\
Rossi, C.~P.\ Silva, and B.~V.\ Carlson,
%Global and consistent analysis of the heavy-ion elastic scattering and fusion processes
Phys.\ Rev.\ C {\bf 69}, 034603 (2004).

\bibitem{cdi07}
A.~I. Chugunov, H.~E. DeWitt, and D.~G. Yakovlev,
%Coulomb tunneling for fusion reactions in dense matter:
%Path integral Monte Carlo versus mean field,
Phys. Rev. D {\bf 76}, 025028 (2007).

%\bibitem{hpy07}
%P.~Haensel, A.~Y.\ Potekhin, and D.~G.\ Yakovlev, {\em Neutron Stars. 1.
%Equation of State and Structure} (Springer, New York, 2007).

\bibitem{ygw06}
D.~G.~Yakovlev, L.~Gasques, and M.~Wiescher, Mon.\ Not.\ Roy.\
Astron.\ Soc.\ {\bf 371}, 1322 (2006).

\bibitem{pc12}
A.~Y.~Potekhin and G.~Chabrier, Astron.\ Astrophys.\ {\bf 538}, A115
(2012).



\end{thebibliography}
\end{document}